\RequirePackage[2020-02-02]{latexrelease}

\documentclass[twocolumn,preprintnumbers]{revtex4}%
\usepackage{amssymb}
\usepackage{amsmath}
\usepackage{graphicx}
\usepackage{epstopdf}
\usepackage{dcolumn}
\usepackage{bm}
\usepackage{xcolor}
\usepackage{ifthen}
\usepackage[normalem]{ulem}
\usepackage{amsfonts}%
\RequirePackage[2020-02-02]{latexrelease}
\graphicspath{{c:/Users/Eyal/MyDocuments/latex/Texfiles/Diamond/dipolarBA/EntangSupp/SDIP/}
	{c:/Users/Eyal/MyDocuments/latex/Texfiles/Diamond/dipolarBA/EntangSupp/SDH/}}
\UseRawInputEncoding
\providecommand{\U}[1]{\protect\rule{.1in}{.1in}}
\providecommand{\U}[1]{\protect\rule{.1in}{.1in}}
\def\showal{1}
\newcommand{\al}[1]{\ifthenelse{\showal=1}{\textcolor{orange}{[[#1]]}}{}}

\newcommand{\eb}[1]{\ifthenelse{\showal=1}{\textcolor{cyan}{[[#1]]}}{}}
\begin{document}
\title{Disentanglement--induced superconductivity}
\author{Eyal Buks}
\email{eyal@ee.technion.ac.il}
\affiliation{Andrew and Erna Viterbi Department of Electrical Engineering, Technion, Haifa
32000, Israel}
\date{\today }

\begin{abstract}
The current study is motivated by a difficulty to reconcile between particle
number conservation and superconductivity. An alternative modeling, which is
based on the hypothesis that disentanglement spontaneously ocuurs in quantum
systems, is explored. The Fermi--Hubbard mode is employed to demonstrate a
disentanglement--induced quantum phase transition into a state having a finite
superconducting order parameter. Moreover, the effect of disentanglement on
Josephson junction's current phase relation is explored.

\end{abstract}
\maketitle





\textbf{Introduction} -- In the Bardeen, Cooper and Schrieffer (BCS) model
\cite{Bardeen_1175}, the Hamiltonian $\mathcal{H}_{\mathrm{BCS}}$ of electrons
in a superconducting metal contains interaction terms proportional to the
operators $B_{\mathbf{k}^{\prime\prime}}^{\dag}B_{\mathbf{k}^{\prime}}^{{}}$,
where $B_{\mathbf{k}^{\prime}}=a_{-\mathbf{k}^{\prime},\downarrow
}a_{\mathbf{k}^{\prime},\uparrow}$ is a pair annihilation operator, and
$a_{\mathbf{k}^{\prime},\sigma}$ annihilates a single particle Fermionic state
having momentum $\hbar\mathbf{k}^{\prime}$ and spin state $\sigma\in\left\{
\uparrow,\downarrow\right\}  $. The operator $B_{\mathbf{k}^{\prime\prime}%
}^{\dag}B_{\mathbf{k}^{\prime}}$ can be expressed as $B_{\mathbf{k}%
^{\prime\prime}}^{\dag}B_{\mathbf{k}^{\prime}}=C_{\mathbf{k}^{\prime
},\mathbf{k}^{\prime\prime}}+B_{\mathbf{k}^{\prime\prime}}^{\dag}\left\langle
B_{\mathbf{k}^{\prime}}^{{}}\right\rangle +\left\langle B_{\mathbf{k}%
^{\prime\prime}}^{\dag}\right\rangle B_{\mathbf{k}^{\prime}}^{{}}-\left\langle
B_{\mathbf{k}^{\prime\prime}}^{\dag}\right\rangle \left\langle B_{\mathbf{k}%
^{\prime}}^{{}}\right\rangle $, where $\left\langle B_{\mathbf{k}^{\prime}%
}^{{}}\right\rangle $ is the expectation value of $B_{\mathbf{k}^{\prime}}%
^{{}}$ in thermal equilibrium, and $C_{\mathbf{k}^{\prime},\mathbf{k}%
^{\prime\prime}}=\left(  B_{\mathbf{k}^{\prime\prime}}^{\dag}-\left\langle
B_{\mathbf{k}^{\prime\prime}}^{\dag}\right\rangle \right)  \left(
B_{\mathbf{k}^{\prime}}^{{}}-\left\langle B_{\mathbf{k}^{\prime}}^{{}%
}\right\rangle \right)  $. In the mean field approximation (MFA) the term
$C_{\mathbf{k}^{\prime},\mathbf{k}^{\prime\prime}}$ is disregarded [see Eq. (18.307) of Ref. \cite{Buks_QMLN}]. This
approximation leads to a mean field Hamiltonian $\mathcal{H}_{\mathrm{MF}}$,
which can be analytically diagonalized by implementing a Bogoliubov transformation.

The MFA greatly simplifies the many--body problem under study, however it
yields some predictions that are arguably inconsistent with what is expected
from the original Hamiltonian $\mathcal{H}_{\mathrm{BCS}}$. Particle number is
conserved by $\mathcal{H}_{\mathrm{BCS}}$, and consequently, it is expected
that in steady state $\left\langle B_{\mathbf{k}^{\prime}}\right\rangle =0$.
In contrast $\left\vert \left\langle B_{\mathbf{k}^{\prime}}\right\rangle
\right\vert $, which is proportional to the BCS energy gap, can become finite
in the MFA. Moreover, the ground state of the mean filed Hamiltonian
$\mathcal{H}_{\mathrm{MF}}$ is continuously degenerate, whereas the ground
state of the BCS Hamiltonian $\mathcal{H}_{\mathrm{BCS}}$ is generically
non--degenerate. The question of MFA validity is related to the spontaneous
symmetry breaking in the Higgs mechanism \cite{Mannheim_1506_04120}.

It was pointed out that the MFA can be, at least partially, justified in the
thermodynamical limit. Particle number conservation implies that $\left\langle
N_{\mathrm{P}}^{2}\right\rangle -\left\langle N_{\mathrm{P}}\right\rangle
^{2}=0$ in steady states, where $N_{\mathrm{P}}=\left(  1/2\right)
\sum_{\mathbf{k}^{\prime}}a_{\mathbf{k}^{\prime},\uparrow}^{\dag}%
a_{\mathbf{k}^{\prime},\uparrow}^{{}}+a_{\mathbf{k}^{\prime},\downarrow}%
^{\dag}a_{\mathbf{k}^{\prime},\downarrow}^{{}}$ is the pair number operator.
In general, the MFA allows the violation of this conservation law (i.e. it
allows non--zero values of $\left\langle N_{\mathrm{P}}^{2}\right\rangle
-\left\langle N_{\mathrm{P}}\right\rangle ^{2}$ in steady state). However, it
was shown that in the MFA both $\left\langle N_{\mathrm{P}}\right\rangle $ and
$\left\langle N_{\mathrm{P}}^{2}\right\rangle -\left\langle N_{\mathrm{P}%
}\right\rangle ^{2}$ are proportional to the volume of the system
\cite{de2018superconductivity}, and thus, the violation of particle number
conservation becomes insignificant in the thermodynamical limit. The mean
field approach has been supported in Ref. \cite{Zyuzin_2308_04508} by showing
that the Ginzburg--Levanyuk parameter is typically small for electrons in
metals. Moreover, it was argued in Ref. \cite{coleman2015introduction} that
the BCS interaction between pairs has an infinite range, and consequently
exact solution of the BCS Hamiltonian $\mathcal{H}_{\mathrm{BCS}}$ can be
derived using a MFA. It was shown in Ref. \cite{Kuzemsky_1530010} that the
Bogoliubov inequality, together with a variational calculation and some
assumptions, can lead to the MFA Hamiltonian $\mathcal{H}_{\mathrm{MF}}$.
Another attempt to rigorously derive the MFA Hamiltonian $\mathcal{H}%
_{\mathrm{MF}}$, which is based on Wick's theorem \cite{Gaudin_89}, has been
presented in \cite{kainth2020superconductivity,Kainth_1904_07138}. However,
this derivation employs a relation, which can be derived from Wick's theorem
only for the case of Gaussian states [see Eq. (16.131) of Ref.
\cite{Buks_QMLN}]. In contrast, the thermal equilibrium state that is derived
from the BCS Hamiltonian $\mathcal{H}_{\mathrm{BCS}}$ is generically non--Gaussian.

The current study is motivated by the arguably limited range of validity of
the MFA, and by the difficulty to reconcile between the spontaneous symmetry
breaking occurring in the superconducting state, and particle number
conservation \cite{Lin_4,Lapa_257002,Ortiz_267002}. An alternative approach,
which is based on a recently--proposed hypothesis that disentanglement
spontaneously occurs in quantum systems, is explored. As is shown below, the
conjecture that disentanglement plays a role in superconductivity is
falsifiable, since it yields predictions that are distinguishable from what is
derived from MFA--based models. In the current study the Fermi--Hubbard model
\cite{Bogolyubov_47,Fabian_1137,Plakida_R11997,Robaszkiewicz_6430,Salwen_064511,Claveau_035023,Iemini_075423,Chen_046401,Xu_067403,Ferreira_115145}
is employed to study the effect of disentanglement on both superconducting
order parameter and current--phase relation (CPR) of a weak link
\cite{Golubov_411}.

\textbf{Disentanglement} -- According to the spontaneous disentanglement
hypothesis, time evolution for the reduced density operator $\rho$ is governed
by a modified master equation given by
\cite{Kaplan_055002,Geller_2200156,Grimaudo_033835,Kowalski_167955,Buks_2400036}%
\begin{equation}
\frac{\mathrm{d}\rho}{\mathrm{d}t}=i\hbar^{-1}\left[  \rho,\mathcal{H}\right]
-\Theta\rho-\rho\Theta+2\left\langle \Theta\right\rangle \rho\;, \label{MME}%
\end{equation}
where $\hbar$ is the Planck's constant, $\mathcal{H}^{{}}=\mathcal{H}^{\dag}$
is the Hamiltonian, the operator $\Theta^{{}}=\Theta^{\dag}$ is allowed to
depend on $\rho$, and $\left\langle \Theta\right\rangle =\operatorname{Tr}%
\left(  \Theta\rho\right)  $. The operator $\Theta$ is given by $\Theta
=\gamma_{\mathrm{H}}\mathcal{Q}^{\left(  \mathrm{H}\right)  }+\gamma
_{\mathrm{D}}\mathcal{Q}^{\left(  \mathrm{D}\right)  }$, where both rates
$\gamma_{\mathrm{H}}$ and $\gamma_{\mathrm{D}}$ are positive, and both
operators $\mathcal{Q}^{\left(  \mathrm{H}\right)  }$ and $\mathcal{Q}%
^{\left(  \mathrm{D}\right)  }$\ are Hermitian. The operator $\mathcal{Q}%
^{\left(  \mathrm{H}\right)  }$, which gives rise to thermalization
\cite{Grabert_161,Ottinger_052119}, is given by $\mathcal{Q}^{\left(
\mathrm{H}\right)  }=\beta\mathcal{U}_{\mathrm{H}}$, where $\mathcal{U}%
_{\mathrm{H}}=\mathcal{H}+\beta^{-1}\log\rho$ is the Helmholtz free energy
operator \cite{Jaynes_579}, $\beta=1/\left(  k_{\mathrm{B}}T\right)  $ is the
thermal energy inverse, $k_{\mathrm{B}}$ is the Boltzmann's constant, and $T$
is the temperature.

For the case of a system composed of indistinguishable particles, the
disentanglement operator $\mathcal{Q}^{\left(  \mathrm{D}\right)  }$ is
derived from two--particle interaction (TPI) \cite{Buks_2400248}. The term in
the Hamiltonian $\mathcal{H}$ accounting for TPI is denoted by $\mathcal{V}$.
In a basis that diagonalizes the TPI, the operator $\mathcal{V}$ is expressed
in terms of the operators $N_{j^{\prime}}N_{j^{\prime\prime}}$, where $N_{j}$
is a number operator associated with the $j$'th single--particle state. In
that basis, each term in $\mathcal{V}$ proportional to $N_{j^{\prime}%
}N_{j^{\prime\prime}}$ contributes to $\mathcal{Q}^{\left(  \mathrm{D}\right)
}$ a term proportional to $Q_{j^{\prime},j^{\prime\prime}}\left\langle
Q_{j^{\prime},j^{\prime\prime}}\right\rangle $, where $Q_{j^{\prime}%
,j^{\prime\prime}}=N_{j^{\prime}}N_{j^{\prime\prime}}-\left\langle
N_{j^{\prime}}\right\rangle \left\langle N_{j^{\prime\prime}}\right\rangle $.
The term $Q_{j^{\prime},j^{\prime\prime}}\left\langle Q_{j^{\prime}%
,j^{\prime\prime}}\right\rangle $ gives rise to suppression of $C_{j^{\prime
},j^{\prime\prime}}^{2}$ with a rate proportional to $\gamma_{\mathrm{D}}$,
where the covariance $C_{j^{\prime},j^{\prime\prime}}$ is defined by
$C_{j^{\prime},j^{\prime\prime}}=\left\langle \left(  N_{j^{\prime}%
}-\left\langle N_{j^{\prime}}\right\rangle \right)  \left(  N_{j^{\prime
\prime}}-\left\langle N_{j^{\prime\prime}}\right\rangle \right)  \right\rangle
=\left\langle Q_{j^{\prime},j^{\prime\prime}}\right\rangle $ [see Eq.
(\ref{MME})]. Alternatively, the covariance $C_{j^{\prime},j^{\prime\prime}}%
$\ can be expressed as $C_{j^{\prime},j^{\prime\prime}}=p_{j^{\prime
},j^{\prime\prime}}-p_{j^{\prime}}p_{j^{\prime\prime}}$, where $p_{j}$ is the
probability that state $j$ is occupied, and $p_{j^{\prime},j^{\prime\prime}}$
is the probability that states $j^{\prime}$ and $j^{\prime\prime}$ are both occupied.

\textbf{Fermi-Hubbard model} -- Consider an array of sites occupied by
Fermions. Single site occupation energy, nearest neighbors hopping and TPI are
characterized by the real parameters $\mu$, $t$ and $U$, respectively. The
creation and annihilation operators corresponding to site $l$ with spin state
$\sigma\in\left\{  \uparrow,\downarrow\right\}  $ are denoted by $a_{l,\sigma
}^{\dag}$ and $a_{l,\sigma}^{{}}$, respectively. The operators $a_{l,\sigma
}^{\dag}$ and $a_{l,\sigma}^{{}}$ satisfy Fermionic anti-commutation
relations. The Fermi-Hubbard Hamiltonian $\mathcal{H}$ is given by
$\mathcal{H}=\mathcal{H}_{0}+\mathcal{V}$, where the single--particle part
$\mathcal{H}_{0}$ is%
\begin{align}
\mathcal{H}_{0}  &  =-t\sum_{\sigma\in\left\{  \uparrow,\downarrow\right\}
}\sum_{\left\langle l^{\prime},l^{\prime\prime}\right\rangle }\left(
a_{l^{\prime},\sigma}^{\dag}a_{l^{\prime\prime},\sigma}^{{}}+a_{l^{\prime
\prime},\sigma}^{\dag}a_{l^{\prime},\sigma}^{{}}\right) \nonumber\\
&  -\mu\sum_{\sigma\in\left\{  \uparrow,\downarrow\right\}  }\sum
_{l}a_{l,\sigma}^{\dag}a_{l,\sigma}^{{}}\ ,\nonumber\\
&  \label{H0 HM}%
\end{align}
where $\left\langle l^{\prime},l^{\prime\prime}\right\rangle $ denotes that
$l^{\prime}$ and $l^{\prime\prime}$ are nearest neighbors, the TPI part is
given by%
\begin{equation}
\mathcal{V}=U\sum_{l}\left(  N_{l,\uparrow}-\frac{1}{2}\right)  \left(
N_{l,\downarrow}-\frac{1}{2}\right)  \ , \label{V HM}%
\end{equation}
and the Fermionic number operator $N_{l,\sigma}$ is given by $N_{l,\sigma
}=a_{l,\sigma}^{\dag}a_{l,\sigma}^{{}}$.

The term $N_{l,\uparrow}N_{l,\downarrow}$ in the TPI part $\mathcal{V}$ [see
Eq. (\ref{V HM})] can be expressed as $N_{l,\uparrow}N_{l,\downarrow
}=\mathcal{C}_{l}+N_{l,\uparrow}\left\langle N_{l,\downarrow}\right\rangle
+\left\langle N_{l,\uparrow}\right\rangle N_{l,\downarrow}-\left\langle
N_{l,\uparrow}\right\rangle \left\langle N_{l,\downarrow}\right\rangle $,
where $\mathcal{C}_{l}=\left(  N_{l,\uparrow}-\left\langle N_{l,\uparrow
}\right\rangle \right)  \left(  N_{l,\downarrow}-\left\langle N_{l,\downarrow
}\right\rangle \right)  $. In the MFA, i.e. when the term $\mathcal{C}_{l}$ is
disregarded, it is well known that the Fermi-Hubbard model supports a
superconducting phase for particular realizations \cite{Arovas_239}.

As was discussed above, disentanglement gives rise to the suppression of the
covariance $\left\langle \mathcal{C}_{l}\right\rangle $. In the rapid
disentanglement approximation \cite{Buks_2400587}, it is assumed that the rate
of disentanglement $\gamma_{\mathrm{D}}$ is sufficiently large to allow
disregarding the term $\mathcal{C}_{l}$. In this limit, the
disentanglement--based model yields predictions that are identical to what is
derived from the standard (i.e. without disentanglement) Fermi-Hubbard model,
when the MFA is implemented, and thus, the disentanglement--based model in
this limit can account for superconductivity, in the same way that the mean
field Fermi--Hubbard model can.

In the current study, the effect of disentanglement is explored, without
assuming that $\gamma_{\mathrm{D}}$ is sufficiently large to validate the
rapid disentanglement approximation. As is demonstrated below, for some cases,
analytical results can be derived from the modified master equation
(\ref{MME}), provided that the size of the under study system is kept
sufficiently small. However, since the rapid disentanglement approximation is
not implemented, analysis commonly becomes intractable in the macroscopic limit.

For the relatively simple systems to be discussed below, it is assumed that
the Fermi-Hubbard array is one dimensional, the number of sites, which is
denoted by $L$, is finite, and the array has a ring configuration, thus, the
last ($l=L$) hopping term $a_{l,\sigma}^{\dag}a_{l+1,\sigma}^{{}%
}+a_{l+1,\sigma}^{\dag}a_{l,\sigma}^{{}}$ [see Eq. (\ref{H0 HM})] is taken to
be given by $a_{L,\sigma}^{\dag}a_{1,\sigma}^{{}}+a_{1,\sigma}^{\dag
}a_{L,\sigma}^{{}}$.

\textbf{Truncation approximation} -- For some cases, dynamics governed by the
modified master equation (\ref{MME}) can be simplified by implementing a
truncation approximation. In this approximation, the operators $\mathcal{H}$
and $\Theta$ are replaced by $P\mathcal{H}P$ and $P\Theta P$, respectively,
where $P$ is a projection operator. For a two--level truncation approximation,
the projection$\ P$ is expressed as $P=\left\vert \psi_{1}\right\rangle
\left\langle \psi_{1}\right\vert +\left\vert \psi_{2}\right\rangle
\left\langle \psi_{2}\right\vert $, where $\left\vert \psi_{1}\right\rangle
$\ and $\left\vert \psi_{2}\right\rangle $ are two orthonormal state vectors
(i.e. $\left\langle \psi_{1}\right.  \left\vert \psi_{1}\right\rangle
=\left\langle \psi_{2}\right.  \left\vert \psi_{2}\right\rangle =1$ and
$\left\langle \psi_{1}\right.  \left\vert \psi_{2}\right\rangle =0$) . The
density operator $\rho$ for that case is expressed in terms of the real vector
$\mathbf{k}=\left(  k_{x},k_{y},k_{z}\right)  $ as%
\begin{equation}
\rho\dot{=}\frac{1+\boldsymbol{\sigma}\cdot\mathbf{k}}{2}\ ,
\end{equation}
where $\boldsymbol{\sigma}=\left(  \sigma_{x},\sigma_{y},\sigma_{z}\right)  $
is the Pauli matrix vector. Similarly, the Hamiltonian is expressed as
$\hbar^{-1}\mathcal{H}\dot{=}\boldsymbol{\sigma}\cdot\boldsymbol{\omega}$,
where $\boldsymbol{\omega}=\left(  \omega_{x},\omega_{y},\omega_{z}\right)  $
is real. It is assumed that $\mathcal{Q}^{\left(  \mathrm{D}\right)
}=Q\left\langle Q\right\rangle $, where $Q\dot{=}q_{0}+\mathbf{q}%
\cdot\boldsymbol{\sigma}$, and both the number $q_{0}$ and the vector
$\mathbf{q}=\left(  q_{x},q_{y},q_{z}\right)  $ are real.

The entropy operator $-\log\rho$ can be expressed as $-\log\rho\dot{=}%
-\log\sqrt{\left(  1-k^{2}\right)  /4}-\left(  \tanh^{-1}k\right)
\boldsymbol{\sigma}\cdot\mathbf{\hat{k}}$, where $k=\left\vert \mathbf{k}%
\right\vert $ and $\mathbf{\hat{k}}=\mathbf{k/}k$, and the operator $\Theta$
as $\Theta=s_{0}+\boldsymbol{\sigma}\cdot\mathbf{s}$, where $s_{0}%
=\gamma_{\mathrm{H}}\left\langle \log\rho\right\rangle +\gamma_{\mathrm{D}%
}q_{0}\left\langle Q\right\rangle $, $\mathbf{s}=\gamma_{\mathrm{H}}\beta
\hbar\boldsymbol{\omega}+\gamma_{\mathrm{D}}\left\langle Q\right\rangle
\mathbf{q}$, and $\left\langle Q\right\rangle =q_{0}+\mathbf{q}\cdot
\mathbf{k}$ [recall the identity $\left(  \boldsymbol{\sigma}\cdot
\mathbf{u}\right)  \left(  \boldsymbol{\sigma}\cdot\mathbf{v}\right)
=\mathbf{u}\cdot\mathbf{v}+i\boldsymbol{\sigma}\cdot\left(  \mathbf{u}%
\times\mathbf{v}\right)  $, and note that the Pauli matrices are all
trace--less]. The modified master equation (\ref{MME}) yields an equation of
motion for $\mathbf{k}$ given by%
\begin{equation}
\frac{\mathrm{d}\mathbf{k}}{\mathrm{d}t}=-2\left(  \mathbf{k}\times
\boldsymbol{\omega}+\mathbf{s}-\left(  \mathbf{s}\cdot\mathbf{k}\right)
\mathbf{k}\right)  \;.
\end{equation}
Note that, generally $\mathbf{s}$ depends on $\mathbf{k}$, and that the vector
$\mathbf{s}-\left(  \mathbf{s}\cdot\mathbf{k}\right)  \mathbf{k}$ is
orthogonal to $\mathbf{k}$ provided that $k=1$ (i.e. $\rho$ represents a pure
state, for which $\operatorname{Tr}\rho^{2}=1$).

When the Hamiltonian $\mathcal{H}$ is time--independent, steady state
solutions of the modified master equation (\ref{MME}) occur at extremum points
of an effective free energy $\left\langle \mathcal{U}_{\mathrm{e}%
}\right\rangle $, which is given by $\left\langle \mathcal{U}_{\mathrm{e}%
}\right\rangle =\gamma_{\mathrm{H}}^{-1}\beta^{-1}\left\langle \Theta
\right\rangle =\left\langle \mathcal{U}_{\mathrm{H}}\right\rangle +\beta
^{-1}\left(  \gamma_{\mathrm{D}}/\gamma_{\mathrm{H}}\right)  \left\langle
\mathcal{Q}^{\left(  \mathrm{D}\right)  }\right\rangle $. In the truncation
approximation $\beta\left\langle \mathcal{U}_{\mathrm{H}}\right\rangle
=\beta\hbar\boldsymbol{\omega}\cdot\mathbf{k}+\left\langle \log\rho
\right\rangle $, where%
\begin{equation}
\left\langle \log\rho\right\rangle =\frac{1-k}{2}\log\frac{1-k}{2}+\frac
{1+k}{2}\log\frac{1+k}{2}\;,
\end{equation}
and $\left\langle \mathcal{Q}^{\left(  \mathrm{D}\right)  }\right\rangle
=\left\langle Q\right\rangle ^{2}=\left(  q_{0}+\mathbf{q}\cdot\mathbf{k}%
\right)  ^{2}$. For a constant $\boldsymbol{\omega}$, the Helmholtz free
energy $\left\langle \mathcal{U}_{\mathrm{H}}\right\rangle $ is minimized at
the thermal equilibrium point $\mathbf{k}=-\tanh\left(  \beta\hbar
\omega\right)  \boldsymbol{\hat{\omega}}$, where the unit vector
$\boldsymbol{\hat{\omega}}$ is given by
$\boldsymbol{\hat{\omega}}=\boldsymbol{\omega}/\left\vert \mathbf{\omega
}\right\vert $ [note that $\mathrm{d}\left\langle \log\rho\right\rangle
/\mathrm{d}k=\tanh^{-1}k$].

For the under--study Fermi-Hubbard model, and for the case of two sites array
(i.e. $L=2$) and $\mu=0$, a two--level truncation approximation, which is
based on a projection onto the subspace spanned by the floor $\left\vert
\mathrm{f}\right\rangle $ (i.e. ground) and ceiling $\left\vert \mathrm{c}%
\right\rangle $ energy eigenstates, becomes applicable provided that
$\left\vert t/U\right\vert \ll1$ \cite{Buks_2400248}. For the case $\mu=0$,
the floor $\left\vert \mathrm{f}\right\rangle $ and ceiling $\left\vert
\mathrm{c}\right\rangle $ states are given by $\left\vert \mathrm{f}%
\right\rangle =\cos\left(  \alpha\right)  \left\vert \mathrm{X}\right\rangle
+\sin\left(  \alpha\right)  \left\vert \mathrm{Y}\right\rangle $ and
$\left\vert \mathrm{c}\right\rangle =\sin\left(  \alpha\right)  \left\vert
\mathrm{X}\right\rangle -\cos\left(  \alpha\right)  \left\vert \mathrm{Y}%
\right\rangle $, where $\left\vert \mathrm{X}\right\rangle =2^{-1/2}\left(
\left\vert 0011\right\rangle +\left\vert 1100\right\rangle \right)  $,
$\left\vert \mathrm{Y}\right\rangle =2^{-1/2}\left(  \left\vert
0110\right\rangle +\left\vert 1001\right\rangle \right)  $, $\alpha=\left(
1/2\right)  \tan^{-1}\left(  -8t/U\right)  $, and $\left\vert \eta_{4}\eta
_{3}\eta_{2}\eta_{1}\right\rangle $ denotes a normalized state, where
$\eta_{1}=\left\langle N_{1,\uparrow}\right\rangle \in\left\{  0,1\right\}  $,
$\eta_{2}=\left\langle N_{1,\downarrow}\right\rangle \in\left\{  0,1\right\}
$, $\eta_{3}=\left\langle N_{2,\uparrow}\right\rangle \in\left\{  0,1\right\}
$ and $\eta_{4}=\left\langle N_{2,\downarrow}\right\rangle \in\left\{
0,1\right\}  $. Note that the disentanglement expectation value $\left\langle
\mathcal{Q}^{\left(  \mathrm{D}\right)  }\right\rangle $\ with respect to the
state $\left\vert \vartheta\right\rangle \equiv\cos\left(  \vartheta\right)
\left\vert \mathrm{X}\right\rangle +\sin\left(  \vartheta\right)  \left\vert
\mathrm{Y}\right\rangle $, where the angle $\vartheta$ is real, is given by
$\left\langle \mathcal{Q}^{\left(  \mathrm{D}\right)  }\right\rangle =\left(
\gamma_{\mathrm{D}}/8\right)  \cos^{2}\left(  2\vartheta\right)  $. Hence, in
the limit $\left\vert t/U\right\vert \ll1$, for which $\left\vert
\mathrm{f}\right\rangle \simeq\left\vert \mathrm{X}\right\rangle $ and
$\left\vert \mathrm{c}\right\rangle \simeq-\left\vert \mathrm{Y}\right\rangle
$, the combined state $2^{-1/2}\left(  \left\vert \mathrm{f}\right\rangle
-\left\vert \mathrm{c}\right\rangle \right)  \simeq\left\vert \vartheta
=\pi/4\right\rangle $ is nearly fully disentangled.

\begin{figure}[ptb]
\begin{center}
\includegraphics[width=3.0in,keepaspectratio]{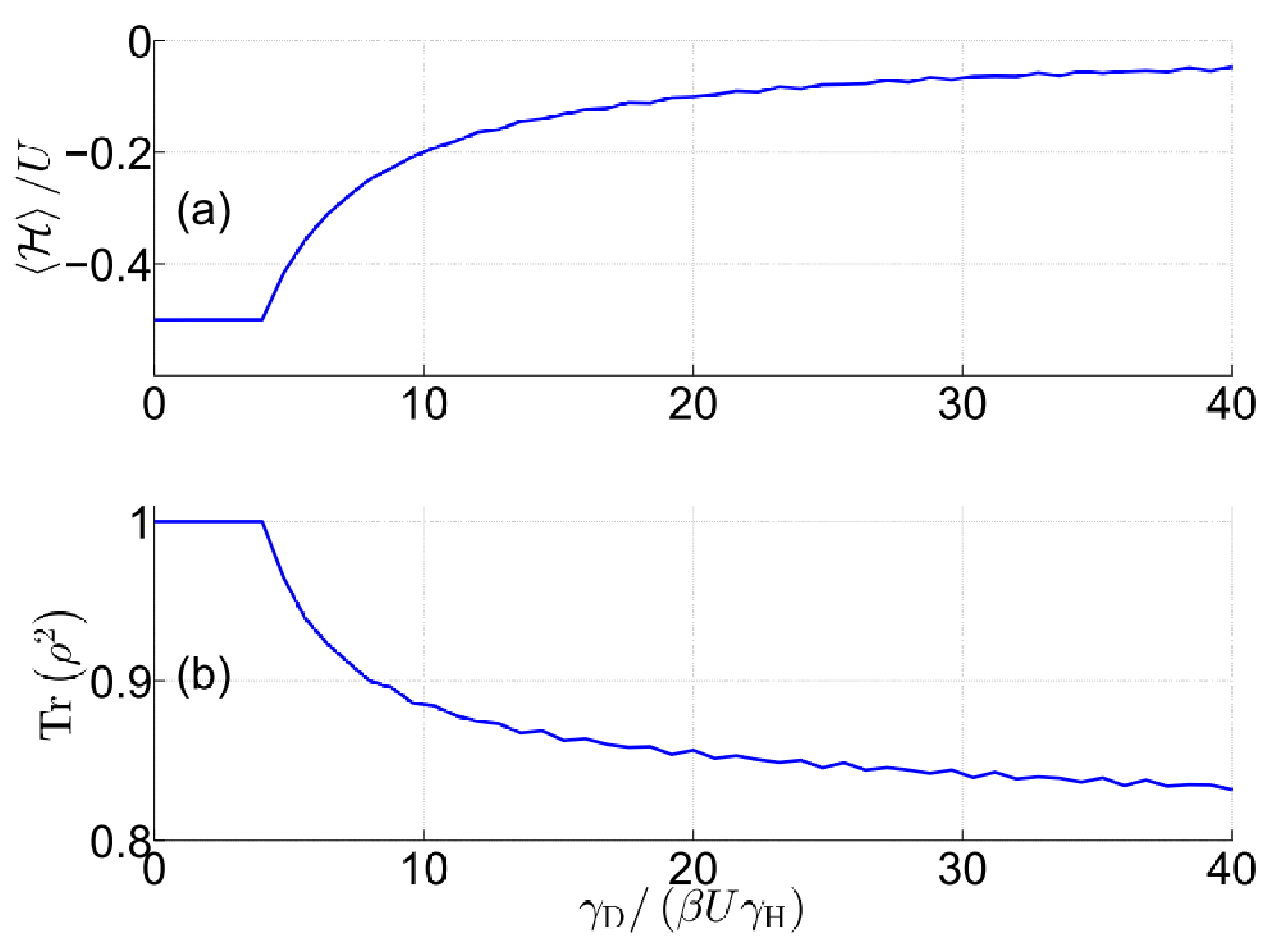}
\end{center}
\caption{{} Fermi-Hubbard model. Steady state values of (a) normalized energy
expectation value $\left\langle \mathcal{H}\right\rangle /U$ and (b) purity
$\operatorname{Tr}\left(  \rho^{2}\right)  $ as a function of the ratio
$\gamma_{\mathrm{D}}/\left(  \beta U\gamma_{\mathrm{H}}\right)  $. A
symmetry--breaking quantum phase transition occurs at $\gamma_{\mathrm{D}%
}/\left(  \beta U\gamma_{\mathrm{H}}\right)  =4$. Assumed parameters' values
are $t/U=10^{-3}$ and $\mu=0$.}%
\label{FigFHMB}%
\end{figure}

The relations $\hbar\boldsymbol{\omega}=E_{0}\left(  0,0,1\right)  $,
$q_{0}=0$ and $\mathbf{q}=\left(  -t/E_{0},0,U/\left(  8E_{0}\right)  \right)
$, where $E_{0}=\left(  1/2\right)  \sqrt{U^{2}+64t^{2}}$, allows analytically
evaluating the effective free energy $\left\langle \mathcal{U}_{\mathrm{e}%
}\right\rangle $. The result reveals that in the low temperature limit, and
for $\left\vert t/U\right\vert \ll1$, a symmetry--breaking quantum phase
transition occurs for this case at $\gamma_{\mathrm{D}}/\left(  \beta
U\gamma_{\mathrm{H}}\right)  =4$. The dependency on the ratio $\gamma
_{\mathrm{D}}/\left(  \beta U\gamma_{\mathrm{H}}\right)  $ of steady state
values of (a) normalized energy expectation value $\left\langle \mathcal{H}%
\right\rangle /U$ and (b) purity $\operatorname{Tr}\left(  \rho^{2}\right)  $
is shown in Fig. \ref{FigFHMB}. The steady state values are calculated by
numerically integrating the modified master equation (\ref{MME}) (without
employing the truncation approximation). The plot in Fig. \ref{FigFHMB}(b)
reveals that the purity $\operatorname{Tr}\left(  \rho^{2}\right)  $ drops
below unity above the phase transition occurring at $\gamma_{\mathrm{D}%
}/\left(  \beta U\gamma_{\mathrm{H}}\right)  =4$.

\begin{figure}[ptb]
\begin{center}
\includegraphics[width=3.0in,keepaspectratio]{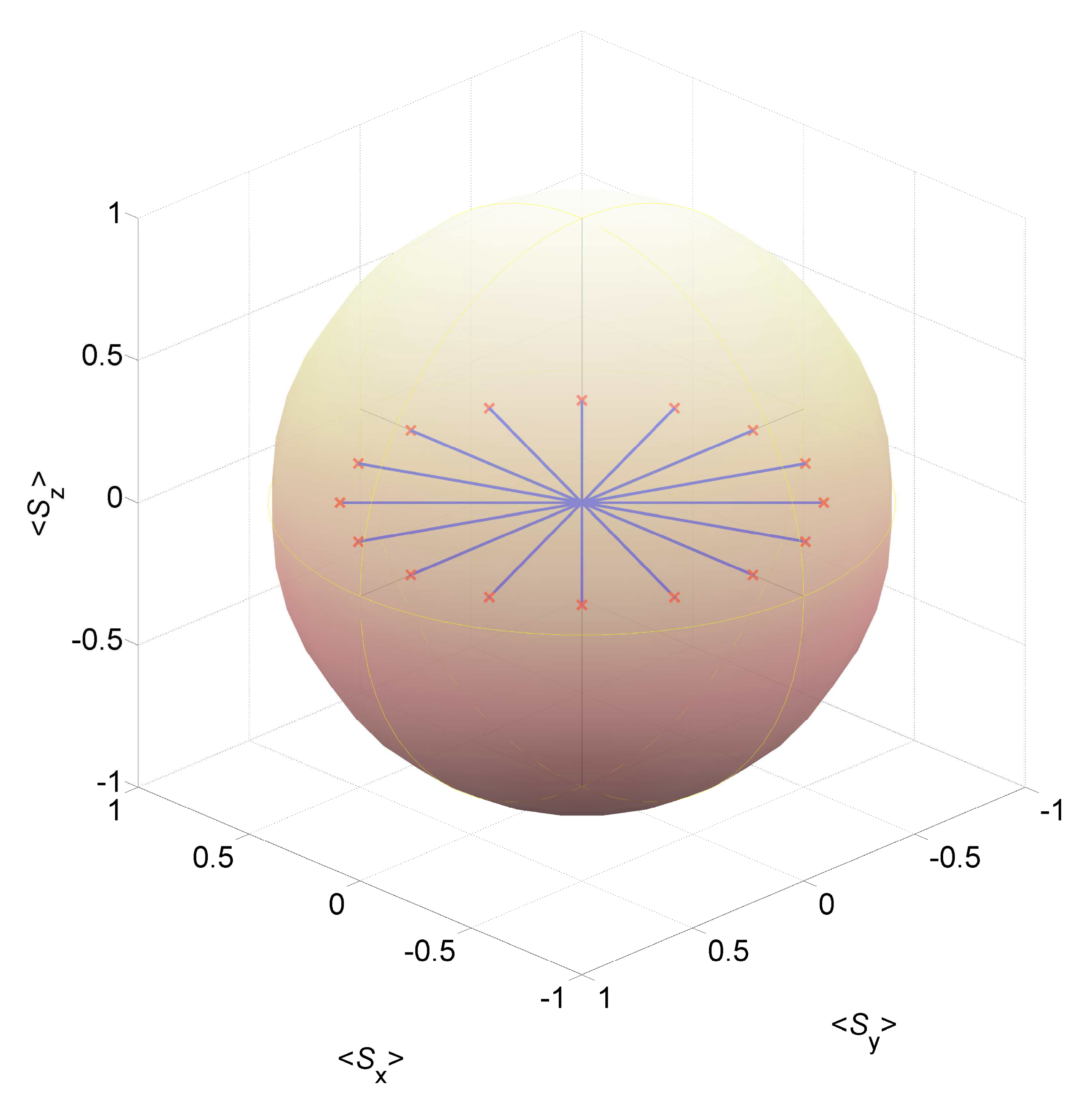}
\end{center}
\caption{{}Disentanglement--induced spontaneous symmetry breaking for the case
$L=2$. Time evolution of the vector $\left\langle \mathbf{S}\right\rangle $
for different initial states located close to the ground state $\left\vert
\mathrm{f}\right\rangle \left\langle \mathrm{f}\right\vert $ [for which
$\left\langle \mathbf{S}\right\rangle =\left(  0,0,0\right)  $]. Assumed
parameters' values are $\epsilon_{\mathrm{s}}=10^{-4}$, $t/U=0.01$, $\mu/U=0$,
and $\gamma_{\mathrm{D}}/(\beta U\gamma_{\mathrm{H}})=50$.}%
\label{FigTSHF}%
\end{figure}

\textbf{Order parameter} -- The plot in Fig. \ref{FigTSHF} demonstrates time
evolution of the vector $\left\langle \mathbf{S}\right\rangle =\left(
\left\langle S_{x}\right\rangle ,\left\langle S_{y}\right\rangle ,\left\langle
S_{z}\right\rangle \right)  $ for the case $L=2$ [the truncation approximation
is not being employed for the numerical integration of the modified master
equation (\ref{MME})]. The vector operator $\mathbf{S}$ is given by
$\mathbf{S}=\sum_{l=1}^{L}\mathbf{S}_{l}$, where $\mathbf{S}_{l}=\left(
S_{l,x},S_{l,y},S_{l,z}\right)  =\Theta_{l}^{\dag}\boldsymbol{\sigma}\Theta
_{l}^{{}}$, and where $\Theta_{l}^{\dag}=\left(
\begin{array}
[c]{cc}%
a_{l,\uparrow}^{\dag}, & a_{l,\downarrow}^{{}}%
\end{array}
\right)  $. The following holds $\left[  S_{l^{\prime},i},S_{l^{\prime\prime
},j}\right]  _{-}=2i\epsilon_{ijk}\delta_{l^{\prime},l^{\prime\prime}%
}S_{l^{\prime},k}$, $S_{l,+}\equiv S_{l,x}+iS_{l,y}=2\mathcal{B}_{l}^{\dag}$,
$S_{l,-}\equiv S_{l,x}-iS_{l,y}=2\mathcal{B}_{l}^{{}}$\ and $S_{l,z}=-1+N_{l}%
$, where $\mathcal{B}_{l}^{{}}=a_{l,\downarrow}^{{}}a_{l,\uparrow}^{{}}$ and
where $N_{l}=N_{l,\uparrow}+N_{l,\downarrow}$, and thus $\mathbf{S}%
_{l^{\prime}}\cdot\mathbf{S}_{l^{\prime\prime}}=2\left(  \mathcal{B}%
_{l^{\prime}}^{\dag}\mathcal{B}_{l^{\prime\prime}}^{{}}+\mathcal{B}%
_{l^{\prime\prime}}^{\dag}\mathcal{B}_{l^{\prime}}^{{}}\right)  +2\left(
1-N_{l^{\prime}}\right)  \delta_{l^{\prime},l^{\prime\prime}}+\left(
1-N_{l^{\prime}}\right)  \left(  1-N_{l^{\prime\prime}}\right)  $ (note that
$\mathcal{B}_{l}^{\dag}\mathcal{B}_{l}^{{}}=a_{l,\uparrow}^{\dag
}a_{l,\downarrow}^{\dag}a_{l,\downarrow}^{{}}a_{l,\uparrow}^{{}}%
=N_{l,\uparrow}^{{}}N_{l,\downarrow}^{{}}$). The variable $\left\langle
S_{x}\right\rangle ^{2}+\left\langle S_{y}\right\rangle ^{2}$ represents an
order parameter.

In the low--temperature limit, and in the absence of disentanglement (i.e. for
$\gamma_{\mathrm{D}}=0$), the ground state density operator $\left\vert
\mathrm{f}\right\rangle \left\langle \mathrm{f}\right\vert $ is a steady state
solution of the modified master equation (\ref{MME}). Note that $\left\langle
\mathbf{S}\right\rangle =\left(  0,0,0\right)  $ for the ground state
$\left\vert \mathrm{f}\right\rangle \left\langle \mathrm{f}\right\vert $.
Above the disentanglement--induced quantum phase transition, i.e. for
$\gamma_{\mathrm{D}}/(\beta U\gamma_{\mathrm{H}})>4$, the ground state becomes
unstable. For the assumed parameters' values used to generate the plot in Fig.
\ref{FigTSHF}, the ratio $\gamma_{\mathrm{D}}/(\beta U\gamma_{\mathrm{H}})$ is
$50$ (see figure caption). The plot shows time evolution for $16$ different
initial pure states, denoted by $\rho_{\mathrm{i}}\left(  \theta_{\mathrm{s}%
}\right)  =\left\vert \psi_{\mathrm{i}}\right\rangle \left\langle
\psi_{\mathrm{i}}\right\vert /\left\langle \psi_{\mathrm{i}}\right.
\left\vert \psi_{\mathrm{i}}\right\rangle $, where $\left\vert \psi
_{\mathrm{i}}\right\rangle $ is given by $\left\vert \psi_{\mathrm{i}%
}\right\rangle =\left\vert \mathrm{f}\right\rangle +\epsilon_{\mathrm{s}%
}\left(  \left\vert 0011\right\rangle +e^{-i\theta_{\mathrm{s}}}\left\vert
1100\right\rangle \right)  $, where $\epsilon_{\mathrm{s}}\ll1$ [i.e.
$\rho_{\mathrm{i}}\left(  \theta_{\mathrm{s}}\right)  \simeq\left\vert
\mathrm{f}\right\rangle \left\langle \mathrm{f}\right\vert $]. Time evolution,
which is obtained by numerically integrating the modified master equation
(\ref{MME}), is shown for $16$ equally--spaced values for the angle
$\theta_{\mathrm{s}}$ in the range $\left[  0,2\pi\right)  $. The plot
demonstrates that the steady state value of $\left\langle \mathbf{S}%
\right\rangle $ (labelled in Fig. \ref{FigTSHF} by red $\times$ symbols) that
is obtained with the initial state $\rho_{\mathrm{i}}\left(  \theta
_{\mathrm{s}}\right)  $ is parallel to the unit vector $\left(  \cos
\theta_{\mathrm{s}},\sin\theta_{\mathrm{s}},0\right)  $. Thus, for this
one--dimensional model, a disentanglement--induced spontaneous symmetry
breaking, which occurs for $\gamma_{\mathrm{D}}/(\beta U\gamma_{\mathrm{H}%
})>4$, gives rise to finite values of the order parameter $\left\langle
S_{x}\right\rangle ^{2}+\left\langle S_{y}\right\rangle ^{2}$.

\textbf{CPR} -- For the case where the one--dimensional array is occupied by
\textit{spinless} Fermions, the Hamiltonian $\mathcal{H}$ is expressed as%
\begin{align}
\mathcal{H}  &  =\sum_{l=1}^{L}\left[  -t_{l}\left(  e^{i\varphi_{l}}%
a_{l}^{\dag}a_{l+1}^{{}}+e^{-i\varphi_{l}}a_{l+1}^{\dag}a_{l}^{{}}\right)
+g_{l}B_{l}^{\dag}B_{l}^{{}}\right] \nonumber\\
&  -\mu\sum_{l=1}^{L}\left(  a_{l}^{\dag}a_{l}^{{}}-\frac{1}{2}\right)
\ .\nonumber\\
&  \label{H no MFA}%
\end{align}
The Fermionic creation and annihilation operators corresponding to site
$l\in\left\{  1,2,\cdots,L\right\}  $ are denoted by $a_{l}^{\dag}$ and
$a_{l}^{{}}$, respectively, the operator $B_{l}$ is given by $B_{l}%
=a_{l+1}a_{l}$ and $B_{L}=a_{1}a_{L}$. It is assumed that $t_{l}=t_{0}%
\delta_{l,L}+t\left(  1-\delta_{l,L}\right)  $ and $g_{l}=g_{0}\delta
_{l,L}+g\left(  1-\delta_{l,L}\right)  $ (i.e. all nearest neighbor site pairs
except of the pair $\left(  L,1\right)  $ share the same coefficients $t_{l}$
and $g_{l}$). The single site occupation energy $\mu$, hopping amplitudes $t$
and $t_{0}$, the phases $\varphi_{l}$, and the pairing amplitudes $g$ and
$g_{0}$\ are all real constants. For the case of an opened chain, $t_{0}=0$
and $g_{0}=0$, whereas $t_{0}=t$ and $g_{0}=g_{{}}$ for the case of a closed ring.

The term $B_{l}^{\dag}B_{l}^{{}}$ can be expressed as $B_{l}^{\dag}B_{l}^{{}%
}=C_{l}+\left\langle B_{l}^{{}}\right\rangle B_{l}^{\dag}+\left\langle
B_{l}^{\dag}\right\rangle B_{l}^{{}}-\left\langle B_{l}^{\dag}\right\rangle
\left\langle B_{l}^{{}}\right\rangle $, where $C_{l}=\left(  B_{l}^{\dag
}-\left\langle B_{l}^{\dag}\right\rangle \right)  \left(  B_{l}^{{}%
}-\left\langle B_{l}^{{}}\right\rangle \right)  $. In the MFA, for which the
term $C_{l}$ is disregarded, the resultant Hamiltonian, which is denoted by
$\mathcal{H}_{\mathrm{K}}$, describes a Kitaev one-dimensional array
\cite{Kitaev_131}. Note that total number of particles is conserved by
$\mathcal{H}$ [see Eq. (\ref{H no MFA})], whereas only the total number mod 2
is conserved by $\mathcal{H}_{\mathrm{K}}$. In the analysis below, the MFA,
which generally enables violation of number conservation, is not implemented.

Consider the case where a magnetic flux given by
$\phi_{\mathrm{e}}=\nu\phi_{0}$ is externally applied to the ring's hole, where
$\nu$ is real, and $\phi_{0}=hc/e$ is the flux quantum (Planck's constant,
vacuum speed of light, and electronic charge are denoted by $h$, $c$, and $e$,
respectively). The effect of the applied flux is taken into account by setting
the phases $\varphi_{l}$ in the Hamiltonian (\ref{H no MFA}) according to
$\varphi_{l}=0$ for $l\in\left\{  1,2,\cdots,L-1\right\}  $, and $\varphi
_{L}=2\pi\nu$ \cite{Byers_46,Buttiker_365}. The circulating current
$\left\langle I\right\rangle $ is calculated using the relation $\left\langle
I\right\rangle =-c\partial\left\langle \mathcal{H}\right\rangle /\partial
\phi_{\mathrm{e}}$ [see Eq. (18.142) of Ref. \cite{Buks_QMLN}], where the steady state energy expectation value $\left\langle
\mathcal{H}\right\rangle $ is evaluated by numerically integrating the
modified master equation (\ref{MME}). For the current case, the
disentanglement operator $\mathcal{Q}^{\left(  \mathrm{D}\right)  }$ is given
by $\mathcal{Q}^{\left(  \mathrm{D}\right)  }=g_{0}Q_{L,1}\left\langle
Q_{L,1}\right\rangle +g\sum_{l=1}^{L-1}Q_{l,l+1}\left\langle Q_{l,l+1}%
\right\rangle $, where $Q_{l^{\prime},l^{\prime\prime}}=N_{l^{\prime}%
}N_{l^{\prime\prime}}-\left\langle N_{l^{\prime}}\right\rangle \left\langle
N_{l^{\prime\prime}}\right\rangle $ (note that $B_{l}^{\dag}B_{l}^{{}}%
=N_{l}N_{l+1}$ and $B_{L}^{\dag}B_{L}^{{}}=N_{L}N_{1}$, where $N_{l}%
=a_{l}^{\dag}a_{l}^{{}}$).

\begin{figure}[ptb]
\begin{center}
\includegraphics[width=3.0in,keepaspectratio]{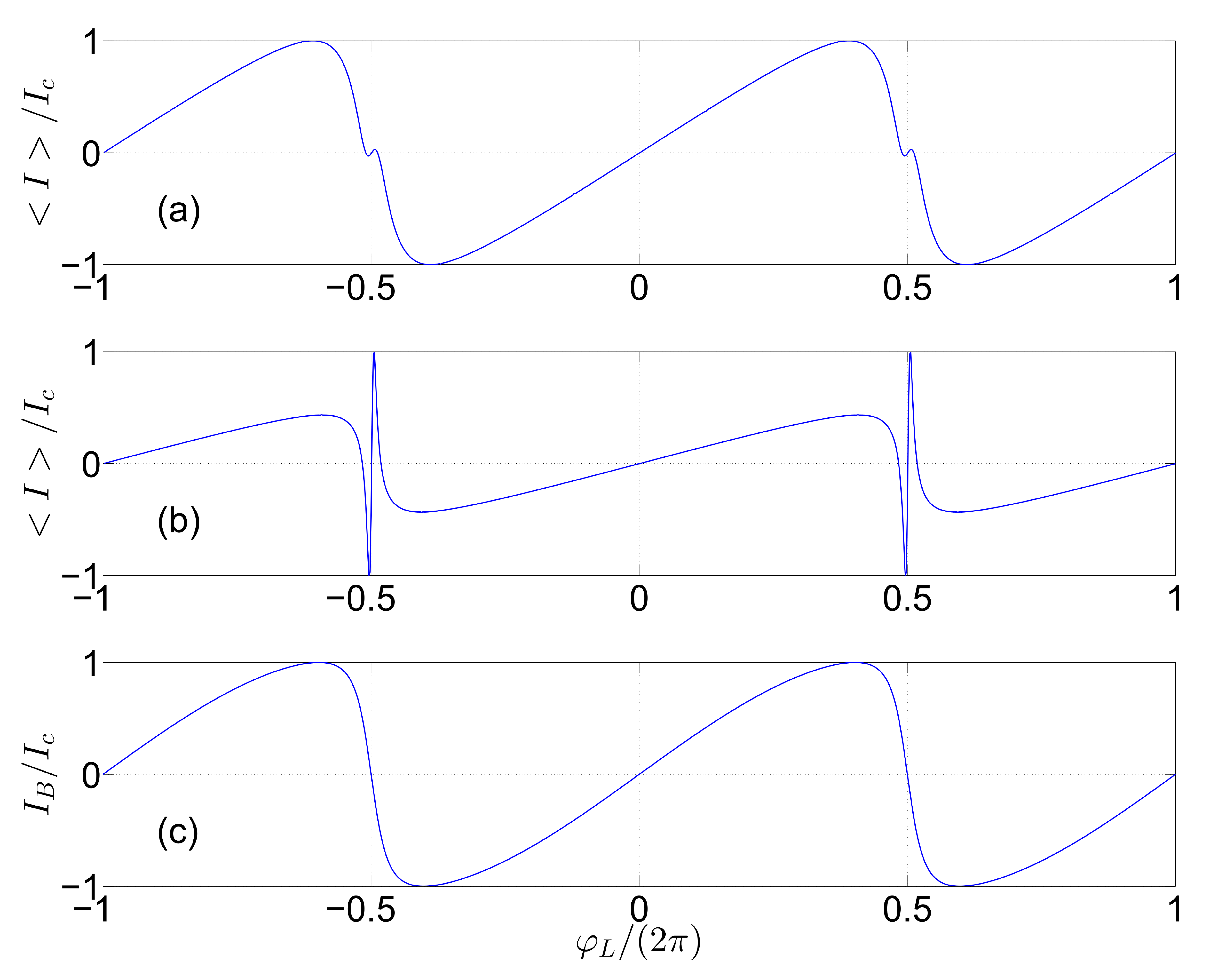}
\end{center}
\caption{CPR. The normalized circulating current $\left\langle I\right\rangle
/I_{\mathrm{c}}$ is shown as a function of normalized applied flux
$\varphi_{L}/\left(  2\pi\right)  =\nu$, where $I_{\mathrm{c}}$ is the
critical current. Assumed parameters' values are, for (a) and (b) $L=5$,
$g/t=1$, $t_{0}/t=0.8$, $g_{0}/t=0$ and $\mu/t=0$, for (a) $\gamma
_{\mathrm{D}}/\gamma_{\mathrm{H}}=5$, for (b) $\gamma_{\mathrm{D}}%
/\gamma_{\mathrm{H}}=10$, and for (c) $\tau=0.99$.}%
\label{FigCPR}%
\end{figure}

The effect of disentanglement on CPR is demonstrated by the plots shown in
Fig. \ref{FigCPR}. The assumed rate of disentanglement $\gamma_{\mathrm{D}}$
for the plots in (a) and (b) is $\gamma_{\mathrm{D}}/\gamma_{\mathrm{H}}=5$,
and $\gamma_{\mathrm{D}}/\gamma_{\mathrm{H}}=10$, respectively. For
comparison, the plot in Fig. \ref{FigCPR}(c) displays the Beenakker--VanHouten
CPR $I_{\mathrm{B}}\left(  \varphi_{L}\right)  $ \cite{Beenakker1991_3056,Beenakker_3836},
which was calculated for a single short channel of transmission $\tau$, and
which is given by $I_{\mathrm{B}}\left(  \varphi_{L}\right)  =I_{\mathrm{c}%
}F\left(  \varphi_{L}\right)  $, where $I_{\mathrm{c}}$ denotes the critical
current, and [see Eq. (A4) of Ref.\cite{Segev_104507}]%
\begin{equation}
F\left(  \varphi_{L}\right)  =\frac{\tau\sin\varphi_{L}}{2\sqrt{2\left(
1-\sqrt{1-\tau}\right)  -\tau}\sqrt{1-\tau\sin^{2}\left(  \varphi
_{L}/2\right)  }}\ .
\end{equation}

The most pronounced effect of disentanglement on the CPR are the sharp
features seen in Fig. \ref{FigCPR}(a) and (b) near half--integer values of the
normalized applied flux $\varphi_{L}/\left(  2\pi\right)  $. These features do
not violate the symmetry relation $I\left(  \varphi_{L}/\left(  2\pi\right)
-n-1/2+x\right)  =-I\left(  \varphi_{L}/\left(  2\pi\right)  -n-1/2-x\right)
$, where $n$ is an integer. Note that some unexplained features obeying the
same symmetry are visible in some spectral measurements of Josephson devices
(e.g. see \cite{Kurter_7130} and Fig. 2 of
\cite{Buks_033807}). Further study is needed to determine whether
disentanglement can account for such experimentally observed features. Note
that a variety of unconventional mechanisms, including topological and multi--band superconductivity, can give rise to CPR having features
that resemble what is seen in Fig. \ref{FigCPR}(a) and (b) (e.g. see Ref.
\cite{Kudriashov_2502_08527}).

\begin{figure}[ptb]
\begin{center}
\includegraphics[width=3.0in,keepaspectratio]{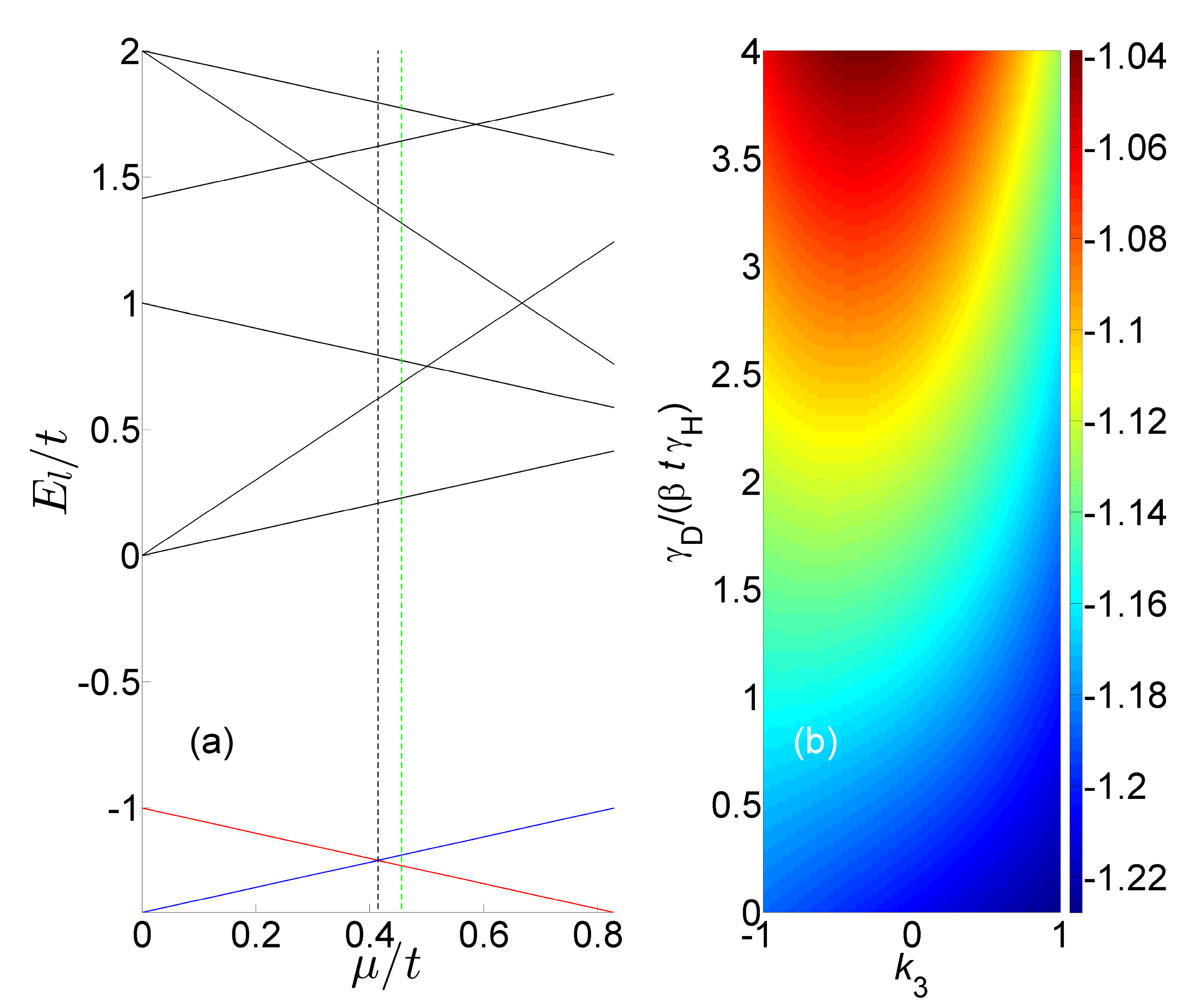}
\end{center}
\caption{Effective free energy. Chain parameters are $L=3$, $g/t=1$,
$\varphi_{l}=0$\ and $t_{0}=g_{0}=0$. (a) The energy eigenvalues $E_{l}$ of
$\mathcal{H}$ (\ref{H no MFA}). (b) The steady state expectation value
$\left\langle \mathcal{U}_{\mathrm{e}}\right\rangle /t $.}%
\label{FigEFE}%
\end{figure}

\textbf{Effective free energy} -- Disentanglement is explored below by
evaluating the effective free energy $\left\langle \mathcal{U}_{\mathrm{e}%
}\right\rangle $ for the spinless one--dimensional array in an open chain
configuration. The energy eigenvalues $E_{l}$ of $\mathcal{H}$ [see Eq.
(\ref{H no MFA})] are shown as a function of $\mu$\ in Fig. \ref{FigEFE}(a),
for the case where $L=3$, $g/t=1$, $\varphi_{l}=0$\ and $t_{0}=g_{0}=0$. For
$\mu<\mu_{\mathrm{c}}$, where $\mu_{\mathrm{c}}=\left(  \sqrt{2}-1\right)  t$
[see the black dashed vertical line in Fig. \ref{FigEFE}(a)], the ground state
is the one--particle state $\left\vert \psi_{1}\right\rangle =2^{-1}\left\vert
100\right\rangle +2^{-1}\left\vert 001\right\rangle +2^{-1/2}\left\vert
010\right\rangle $ [see the blue line in Fig. \ref{FigEFE}(a)], whereas the
two--particle state $\left\vert \psi_{2}\right\rangle =6^{-1/2}\left\vert
110\right\rangle +6^{-1/2}\left\vert 011\right\rangle +2\times6^{-1/2}%
\left\vert 101\right\rangle $ [see the red line in Fig. \ref{FigEFE}(a)]
becomes the ground state for $\mu>\mu_{\mathrm{c}}$.

Consider a reduced density operator $\rho$ having matrix representation in the
basis $\left\{  \left\vert \psi_{1}\right\rangle ,\left\vert \psi
_{2}\right\rangle \right\}  $ given by $\rho\dot{=}\left(  1/2\right)  \left(
1+\mathbf{k}\cdot\boldsymbol{\sigma}\right)  $, where $\mathbf{k}=\left(
k_{1},k_{2},k_{3}\right)  $ is real. The truncated density operator $\rho$ can
be used for approximately calculating the effective free energy $\left\langle
\mathcal{U}_{\mathrm{e}}\right\rangle $ for $\mu\simeq\mu_{\mathrm{c}}$. The
dependency of $\left\langle \mathcal{U}_{\mathrm{e}}\right\rangle $ on $k_{3}$
and $\gamma_{\mathrm{D}}/\left(  \beta t\gamma_{\mathrm{H}}\right)  $ for the
value $\mu/\mu_{\mathrm{c}}=1.1$ [see the green dashed vertical line in Fig.
\ref{FigEFE}(a)] is shown in Fig. \ref{FigEFE}(b) (note that $\left\langle
\mathcal{U}_{\mathrm{e}}\right\rangle $ does not depend on $k_{1}$ and on
$k_{2}$ in the truncation approximation). The color--coded plot of
$\left\langle \mathcal{U}_{\mathrm{e}}\right\rangle $ reveals a
disentanglement--induced transition from monostability to bistability. In the
low temperature limit, and in the absence of disentanglement [i.e. in the
limit $\gamma_{\mathrm{D}}/\left(  \beta t\gamma_{\mathrm{H}}\right)
\rightarrow0$], the effective free energy $\left\langle \mathcal{U}%
_{\mathrm{e}}\right\rangle $ is minimized for the two--particle state
$\left\vert \psi_{2}\right\rangle $. However, for $\gamma_{\mathrm{D}}%
\gtrsim\beta t\gamma_{\mathrm{H}}$, the system becomes bistable [see Fig.
\ref{FigEFE}(b)].

\textbf{Summary} -- Spontaneous disentanglement allows the violation of
particle number conservation, which, in turn, enables a quantum phase
transition induced by symmetry breaking. The Hubbard--Fermi model is employed
for studying the effect of disentanglement on the superconducting order
parameter and on the CPR of a weak link. While the current study is focused on
exploring the effect of disentanglement on small systems, future research will
explore the macroscopic limit using stability analysis \cite{Halperin_521}
(this research direction has been proposed by one of the reviewers of this
paper). Moreover, more realistic theoretical models that can yield
experimentally testable predictions will be developed.

\bibliographystyle{ieeepes}
\bibliography{acompat,Eyal_Bib}

\end{document}